\documentclass{iau}
\usepackage{graphicx}
\usepackage{setspace}
\usepackage{graphics}
\usepackage{graphicx}
\usepackage{amssymb}
\usepackage{longtable}
\usepackage{amsmath}

\title[Formation of discs around SMBH binaries] %% give here short title %%
{Formation of discs around super-massive black hole binaries}

\author[F.G.\ Goicovic et al.]   %% give here short author list %%
{Felipe G.\ Goicovic$^1$,
 Jorge Cuadra$^1$ \and Alberto Sesana$^2$}

\affiliation{$^1$Instituto de Astrof\'isica, Pontificia Universidad Cat\'olica de Chile,  Santiago, Chile \\ 
$^2$ Max-Planck-Institut f\"ur Gravitationsphysik, Potsdam-Golm, Germany\\
email: {\tt fgarrido@astro.puc.cl; jcuadra@astro.puc.cl; alberto.sesana@aei.mpg.de} \\[\affilskip]}

\pubyear{2014}
\volume{312}  %% insert here IAU Symposium No.
\pagerange{000--000}
\jname{Star Clusters and Black Holes in galaxies across cosmic time}
\editors{A.C. Editor, B.D. Editor \& C.E. Editor, eds.}
\begin{document}

\maketitle

\begin{abstract}
We model numerically the evolution of $10^4M_\odot$ turbulent
molecular clouds in near-radial infall onto $10^6M_\odot$, equal-mass
super-massive black hole binaries, using a modified version of
the SPH code \textsc{gadget}-3.  We investigate the different gas structures
formed depending on the relative inclination
between the binary and the cloud orbits.  Our first results indicate
that an aligned orbit produces mini-discs around each black hole,
almost aligned with the binary; a perpendicular orbit produces
misaligned mini-discs; and a counter-aligned orbit produces a
circumbinary, counter-rotating ring.  
%% add here a maximum of 10 keywords, to be taken form the file <Keywords.txt>
\end{abstract}

\firstsection % if your document starts with a section,
              % remove some space above using this command.
\section{Introduction}

Super-massive black holes (SMBHs) are ubiquitous in galactic nuclei (\cite{rich98}), and binaries of these massive objects are a likely product of the hierarchical evolution of structures in the universe. 
After a galaxy merger, where both progenitors host a SMBH, different mechanisms are responsible for the evolution of the binary orbit depending on its separation (see review by \cite{Colpi2014}). Dynamical interaction with stars appears to be efficient to bring the SMBHs down  to parsec scales only, what is known as the ``last parsec problem'' (\cite{Begelman1980,Yu2002}). A possible way to overcome this barrier and merge the SMBHs within a Hubble time is interaction with gas. Many theoretical and numerical studies have focused on the orbital evolution of a sub-parsec binary surrounded by a circumbinary disc (e.g. \cite{ArmNat05,C09,Pau2013,Nix13,Roedig2014}). However, the exact mechanism that would produce such discs is still unclear; it is necessary an efficient transport of gas from thousands or hundreds of parsecs to the central parsec.

Turbulence and gravitational instabilities in the interstellar medium, through the formation of clumps, allow portions of gas to travel almost unaffected by its surrounding, enhancing the probability of reaching the galactic nuclei (\cite{Hobbs2011}). A possible manifestation of these events is the putative molecular cloud that resulted in the unusual distribution of young stars orbiting our Galaxy's SMBH. In particular, the simulation of \cite[Bonnell \& Rice (2008)]{BR08} shows a spherical, turbulent cloud falling with a very low impact parameter ($\sim$0.1 pc) onto a one million solar masses SMBH. 
 
%The interstellar medium is thought to be multiphase where turbulence and gravitational instabilities can trigger the formation of clumps. The mass spectrum and frequency of the clumps suggest that they are likely the seeds for molecular clouds (\cite{Agertz2009}). Several studies, attempting to explain the unusual distribution of stars orbiting our Galaxy's SMBH, have shown that portions of infalling gas, in the form of gas clouds, can be captured by a SMBH to form an eccentric disc that eventually fragments to form stars (e.g. \cite{BR08,HobNay09,Map12}). Even though the characteristics of the clouds are quite different, there is a key parameter present on each of these simulations: the near radial infall of material.  In particular, the simulation of \cite[Bonnell \& Rice (2008)]{BR08} shows a spherical, turbulent cloud falling with a very low impact parameter ($\sim$0.1 pc) onto a one million solar masses SMBH. 

Assuming that these accretion events are common in galactic nuclei, the goal of our work is to model such an event onto a binary instead of a single SMBH.  In particular, we are interested on the properties of the discs that will form given different relative orientations between the orbital angular momenta of the cloud and the binary. Notice that this study is complementary to that shown in \cite{Dunhill2014}, as we are modeling clouds with very low orbital angular momentum.

\section{The numerical model}

We model the interaction between the binaries and clouds using a modified version of the SPH code \textsc{gadget}-3 (\cite{Springel2005}). The cloud is represented using over $4\times 10^6$ gas particles with a total mass of $10^4M_{\odot}$, an initial turbulent velocity field and uniform density. The SMBHs are modelled as two equally-massive sink particles that interact only through gravity and can accrete SPH particles. The total mass of the binary is $10^6M_{\odot}$, and its initial orbit is Keplerian and circular.

\begin{figure}
\centering
 \includegraphics[width=0.6\textwidth]{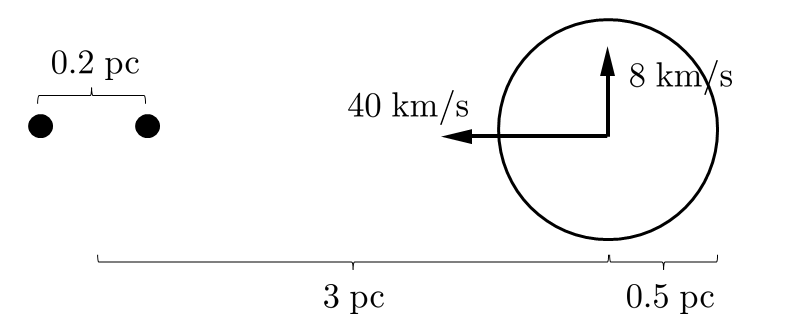}
 \caption{Initial setup of the simulations. The circle on the right represents the cloud, while the black, solid circles are the SMBHs.}
 \label{initial}
\end{figure}

The physical setup of the simulation is shown in Figure \ref{initial}. The initial velocity of the cloud yields a highly eccentric ($e\approx 0.98$), bound orbit where the pericenter distance is $r_{\rm peri}\approx 0.07$ pc, which is less than the binary radius, making the interaction between the gas and SMBHs very strong.
As we expect  clouds approaching the binary from different directions, we model systems with three different inclinations between the cloud and binary orbits: aligned, perpendicular and counter-aligned.

\section{Results: formation and properties of the gaseous discs}

In this section we present the main results of the simulations with the different inclinations, in particular the discs that form around the binary and each SMBHs.

\subsection{Aligned orbits}

\begin{figure}[htbp]
{\includegraphics[width=0.6\textwidth]{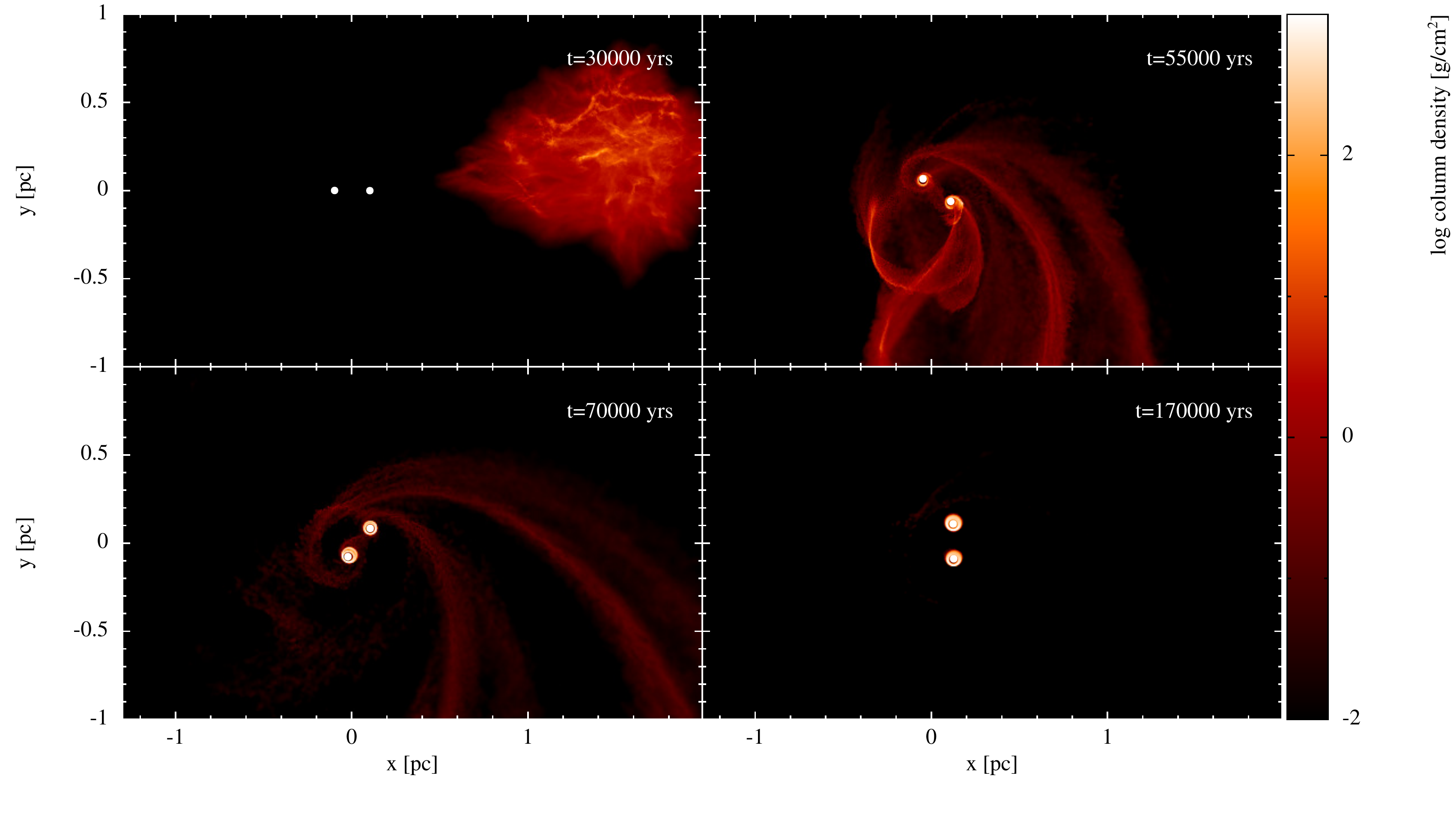}}
{ \includegraphics[width=0.4\textwidth]{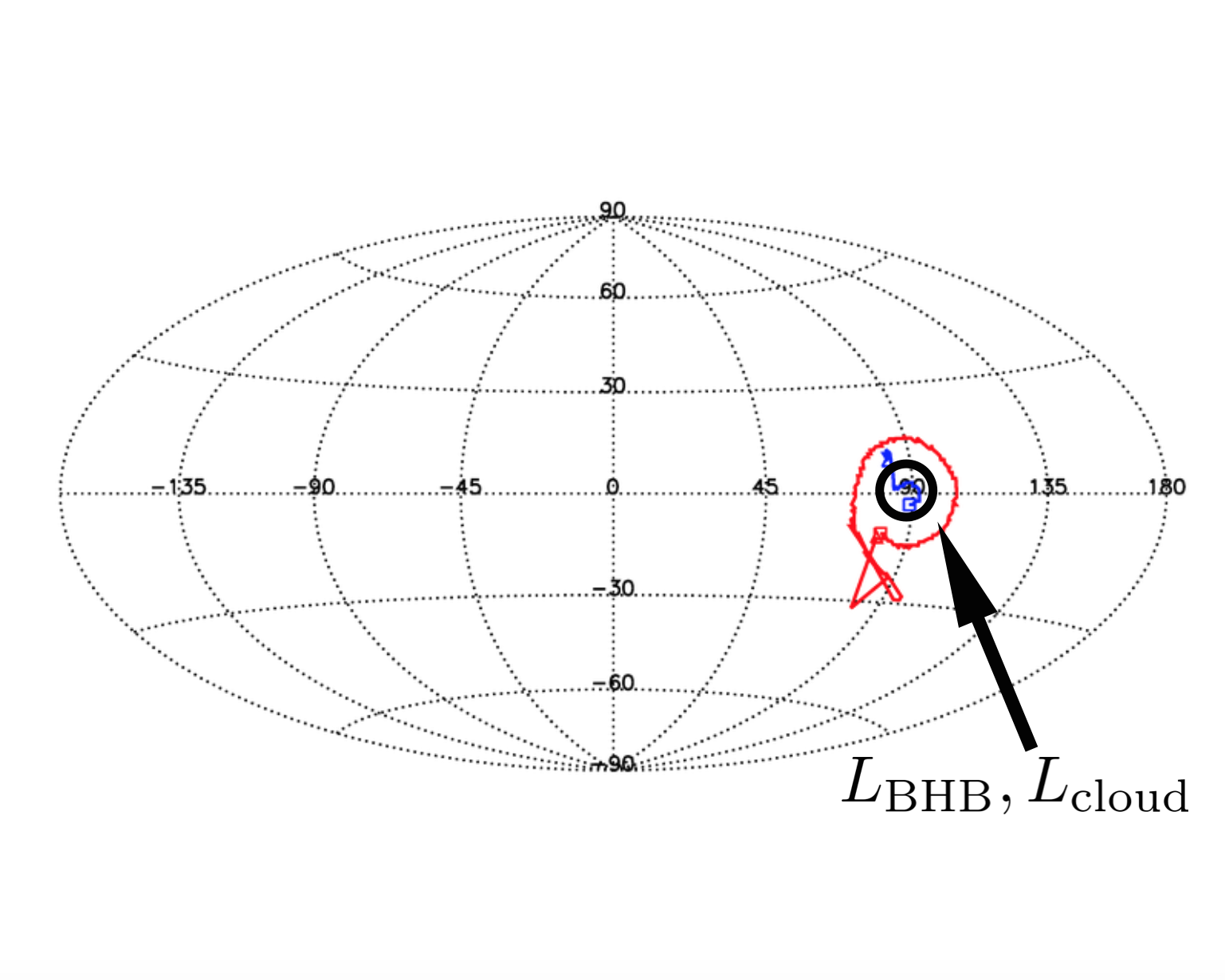}}
 \caption{Left: Column density map of the aligned simulation at different times. For reference, the cloud and the binary move on the x-y plane. (For animations check \url{http://www.astro.puc.cl/~fgarrido/animations})
 Right: Time evolution of the angular momentum direction of the mini-discs formed around each black hole (red and blue lines), in a Hammer projection. The triangle shows the formation of the disc and the square the end of the simulation.}
 \label{BHBxy}
\end{figure}

On the left panel of Figure \ref{BHBxy} we show the column density map of the simulation at different times, where we can see how the interaction develops. As the gas falls almost radially onto the binary, around 80\% of the cloud is accreted by the SMBHs. Most of the remaining gas is pushed away due to an efficient slingshot. The bound material forms a tail that get stretched and diluted over time, feeding ``mini-discs" that form around each SMBH.

To measure the alignment between the binary orbit and the mini-discs, we compute its angular momentum on the corresponding black hole reference frame. We show the time evolution of the direction of both discs on the Hammer projection of Figure \ref{BHBxy}. Here we observe that they tend to align with the orbit of the binary, as expected, although one disc is slightly tilted with respect to the aligned position and also precesses around that position. This behavior could have distinctive electromagnetic signatures. For example, the misalignment could affect the variability of spectral lines, or each disc have different polarisation. The precession could be observed if  jets are launched from the SMBHs and align with the mini-discs.

\subsection{Perpendicular orbits}

\begin{figure}[htbp]
{\includegraphics[width=0.6\textwidth]{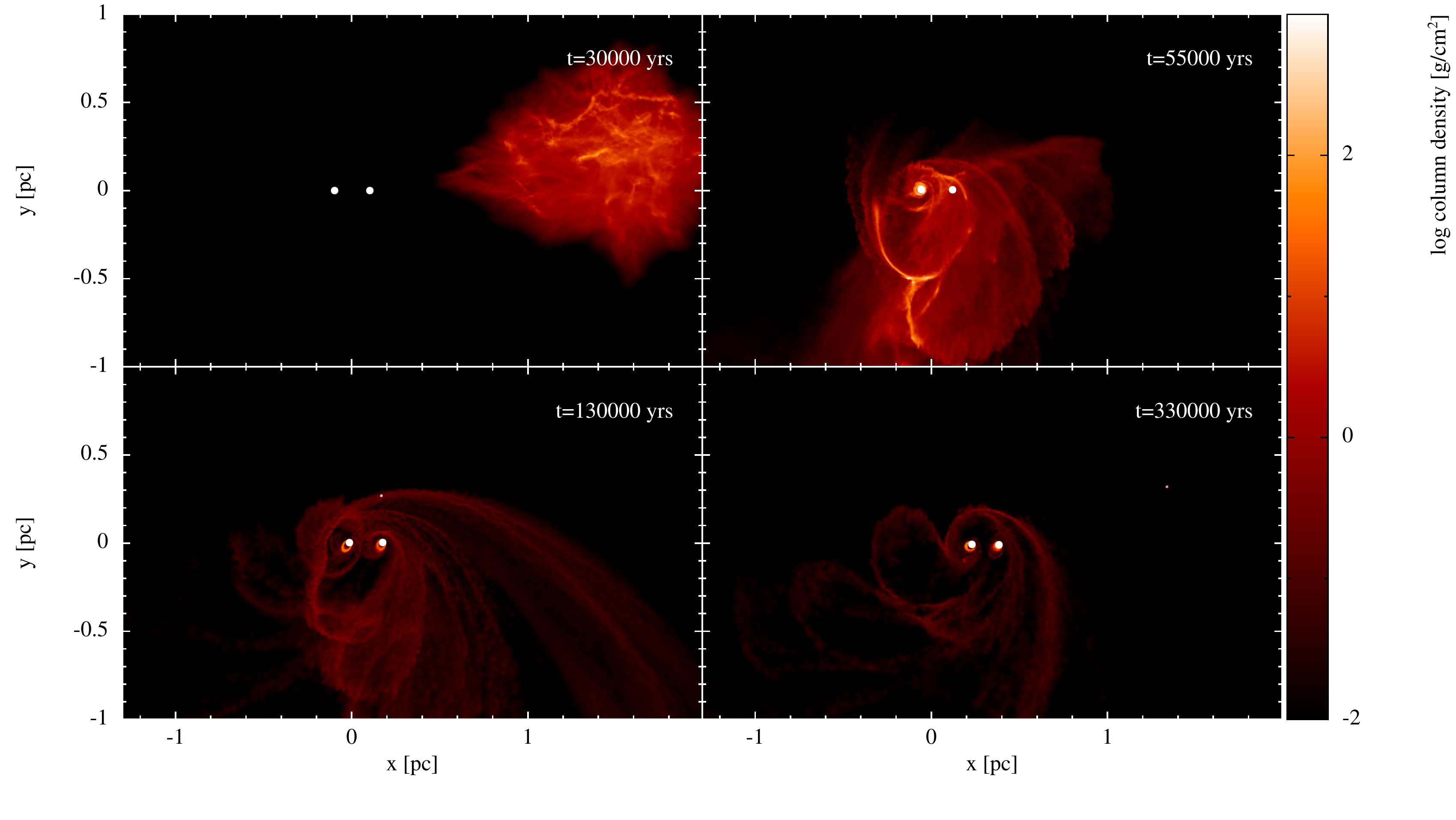}}
 { \includegraphics[width=0.4\textwidth]{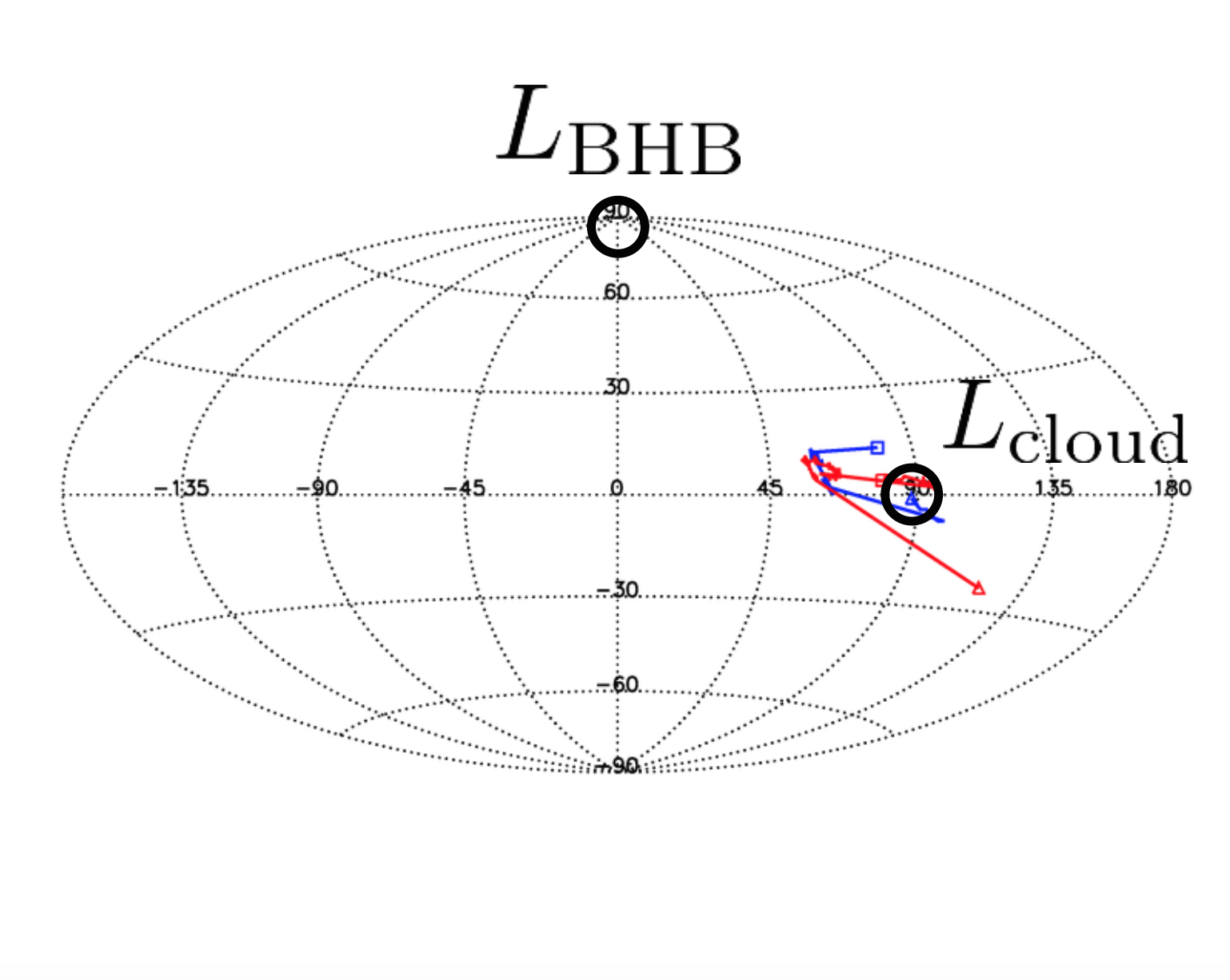}}
 \caption{As Fig.~\ref{BHBxy}, but for the model with perpendicular orbits.  In this case  
 the cloud moves on the x-y plane while the binary is on the x-z plane.}
  \label{BHBxz}
\end{figure}

With this inclination, as in the previous case, around 80\% of the cloud mass is added to the SMBHs. However, the interaction between the gas and the binary, that we can see in Figure \ref{BHBxz}, is completely different respect to the aligned case. Due to a less efficient slingshot, most of the remaining material stays bound to the system and it retains its original angular momentum, forming an unstable structure around the binary. The gas that reaches the SMBHs also produce mini-discs, but they are less massive and more intermittent than in the aligned case. 

The direction of the mini-discs, shown on the right panel of Figure \ref{BHBxz}, shows that they tend to follow the original direction of the cloud, which makes them completely misaligned respect to the binary orbit. As well as the previous case, this could have distinctive signatures on the variability of lines or the direction of possible jets.

\subsection{Counter-aligned orbits}

In this case we have that the interaction of the binary with the gas produces shocks that cancel angular momentum, allowing the SMBHs to accrete even more material than in the previous cases; around 90\% of the cloud is swallowed. The remaining material forms a tail that, due to the gravitational torques, produces a circumbinary ring. In this case we do not observe mini-discs during the entire length of the simulation.

\begin{figure}[htbp]
{\includegraphics[width=0.6\textwidth]{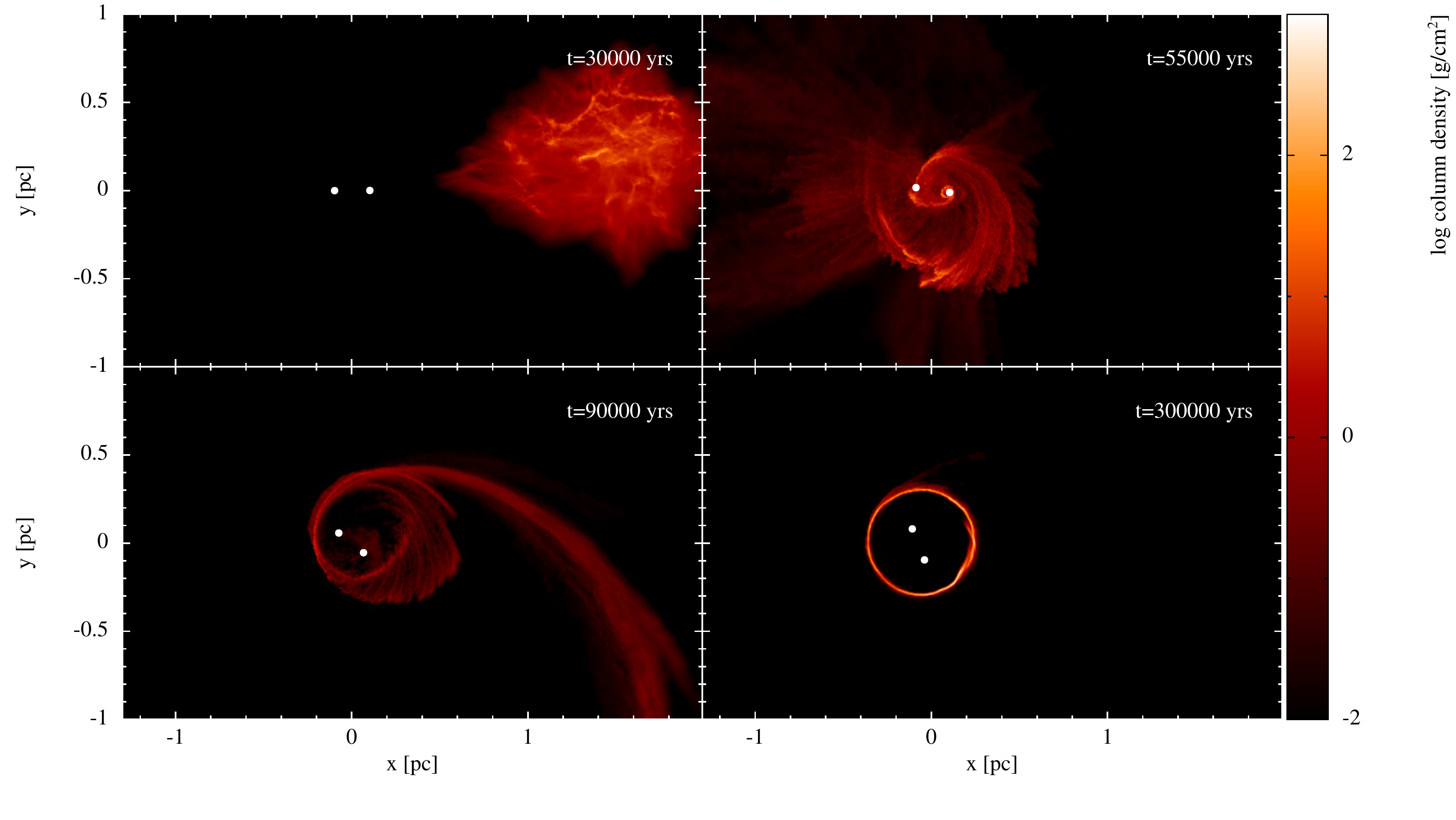}}
{\includegraphics[width=0.42\textwidth]{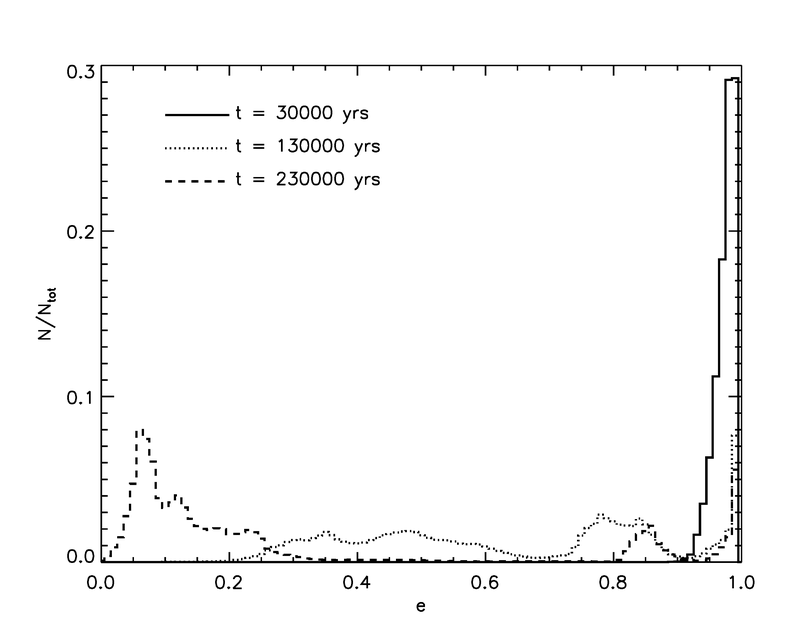}}
 \caption{Left: As Fig.~\ref{BHBxy}, but for the model with the counter-aligned orbits. In this case, the cloud and the binary move on the x-y plane, but the latter is rotating clockwise. 
  Right: Eccentricity distribution of the gas for three different times in the simulation.}
 \label{BHB-xy}
\end{figure}

Finally, we compute the eccentricity distribution of the gas on three different stages, as shown on the right panel of Figure \ref{BHB-xy}. It is interesting that, if there is star formation in the ring, the stars would have very different orbits around the binary depending on the formation time-scale. For a very rapid star formation we could have highly eccentric orbits (solid line), while for a slow process the stars could be distributed on a narrow ring with nearly circular orbits (dashed line). %Detecting these type of differences for stellar orbits in galactic nuclei might be possible with the next generation of telescopes.

\section{Conclusions and outlook}
The preliminary results presented here show that accretion events onto binaries result in very different disc morphologies, depending on the relative inclination between the cloud and the binary orbits.  Our work in progress is extending this study to larger impact parameters, showing lower accretion rates, and exploring the effect of the interaction on the black holes orbit and spin evolution.  These results are likely to have important implications on the multi-messenger future studies of SMBH binaries and on the long-term evolution of these systems.

\subsubsection*{Acknowledgements}
Column density maps were created with {\sc splash} by \cite{Splash}.
We acknowledge support from CONICYT-Chile through PCHA/Doctorado Nacional, FONDECYT (1141175),
Basal (PFB0609), Anillo (ACT1101), Redes (120021) and Exchange (PCCI130064) grants.


\begin{thebibliography}{}

%\bibitem[Agertz \etal\ 2009]{Agertz2009} Agertz, O., Lake, G., 
%Teyssier, R., et al.\ 2009, \textit{MNRAS}, 392, 294

%\bibitem[Alig \etal\ 2011]{Alig2011} Alig, C., Burkert, A., 
%Johansson, P.~H., \& Schartmann, M.\ 2011, \textit{MNRAS}, 412, 469

\bibitem[Amaro-Seoane \etal\ 2013]{Pau2013} Amaro-Seoane, P., 
Brem, P., \& Cuadra, J.\ 2013, \textit{ApJ}, 764, 14

\bibitem[Armitage 
\& Natarajan 2005]{ArmNat05} Armitage, P.~J., \& Natarajan, P.\ 2005, \textit{ApJ}, 634, 921

\bibitem[Begelman \etal\ 1980]{Begelman1980} Begelman, M.~C., 
Blandford, R.~D., \& Rees, M.~J.\ 1980, \textit{Nature}, 287, 307

\bibitem[Bonnell 
\& Rice 2008]{BR08} Bonnell, I.~A., \& Rice, W.~K.~M.\ 2008, \textit{Science}, 321, 1060

\bibitem[Colpi 2014]{Colpi2014} Colpi, M.\ 2014, \textit{SSR}, 183, 189

\bibitem[Cuadra \etal\ 2009]{C09} Cuadra, J., Armitage, 
P.~J., Alexander, R.~D., \& Begelman, M.~C.\ 2009, \textit{MNRAS}, 393, 1423

\bibitem[Dunhill \etal\ (2014)]{Dunhill2014} Dunhill, A.~C., 
Alexander, R.~D., Nixon, C.~J., \& King, A.~R.\ 2014, \textit{MNRAS}, 445, 2285

%\bibitem[Hobbs 
%\& Nayakshin 2009]{HobNay09} Hobbs, A., \& Nayakshin, S.\ 2009, \textit{MNRAS}, 394, 191

\bibitem[Hobbs \etal\ 2011]{Hobbs2011} {Hobbs}, A., {Nayakshin}, S., {Power}, C. \& {King}, A. \ 2011, \textit{MNRAS}, 413, 2633

%\bibitem[Lucas \etal\ 2013]{Luc13} Lucas, W.~E., Bonnell, 
%I.~A., Davies, M.~B., \& Rice, W.~K.~M.\ 2013, \textit{MNRAS}, 433, 353

%\bibitem[Mapelli \etal\ 2012]{Map12} Mapelli, M., Hayfield, 
%T., Mayer, L., \& Wadsley, J.\ 2012, \textit{ApJ}, 749, 168

\bibitem[Nixon \etal\ 2013]{Nix13} Nixon, C., King, A., 
\& Price, D.\ 2013, \textit{MNRAS}, 434, 1946 

\bibitem[Price (2007)]{Splash} Price 
D.~J., 2007, PASA, 24, 159 

\bibitem[Richstone \etal\ 1998]{rich98}
Richstone, D., %Ajhar, E. A., Bender, R., Bower, G., Dressler, A., Faber, S. M., Filippenko, A. V., Gebhardt, K., Green, R., Ho, L. C., Kormendy, J., Lauer, T. R., Magorrian, J. \& Tremaine, S., 
et al.
1998,
\textit{Nature}, 395

\bibitem[Roedig 
\& Sesana 2014]{Roedig2014} Roedig, C., \& Sesana, A.\ 2014, \textit{MNRAS}, 439, 3476

%\bibitem[Sanders 1998]{Sanders98} Sanders, R. H., 1998, \textit{MNRAS}, 294, 35

\bibitem[Springel 2005]{Springel2005} Springel, V.\ 2005, \textit{MNRAS}, 
364, 1105

\bibitem[Yu 2002]{Yu2002} Yu, Q.\ 2002, \textit{MNRAS}, 331, 935 

\end{thebibliography}
\end{document}